\documentclass[ article, aps,]{revtex4-1}
\usepackage{amsmath}

\usepackage{graphicx}
\usepackage{dcolumn}
\usepackage{bm}
\usepackage{color}

\newcommand{\beq}{\begin{equation}}
\newcommand{\enq}{\end{equation}}
\newcommand{\beqa}{\begin{eqnarray}}
\newcommand{\beqast}{\begin{eqnarray*}}
\newcommand{\enqa}{\end{eqnarray}}
\newcommand{\enqast}{\end{eqnarray*}}

\begin{document}

\title{  Forward scattering amplitudes of pp and p\=p  with crossing symmetry and scaling properties }

\author {A. K. Kohara }
\address{CPHT, CNRS, Ecole polytechnique, IP Paris, F-91128 Palaiseau, France}  


\begin{abstract}
We analyse the pp and p\=p elastic scattering amplitudes using the data of several CERN and FERMILAB experiments, revisiting ideas proposed by Andr\'e Martin based on analytic continuation and crossing symmetry.  Introducing a new form for the scaling function together with the analytical forms from COMPETE at $t=0$ we show that the data are consistent with the crossing symmetry of the scattering amplitudes from $\sqrt{s}= $ 23 GeV to 13 TeV for $-t\leq 0.2$ GeV$^{2}$. Analiticity and crossing symmetry automatically satisfy the  dispersion relations and their derivatives. The real part reproduces the  zero predicted by Martin, which is crucial to describe with precision the differential cross section  in the forward range at high energies. Since the free parameters of the model are energy independent, the analytical form of the amplitude allows predictions  for intermediate and higher energies.
\end{abstract}


         \maketitle

\section{Introduction of the model}
Theoretical and phenomenological approaches for the description of  pp and p\=p elastic scattering aim to determine the dynamics and kinematical dependence of the amplitudes,  described in terms of the two variables $s$ and $t$. 
In Regge theory the rise of the hadronic total cross section can be described by  the Pomeron trajectory  linear in $t$, with power dependence on $s$ in the case of a simple pole. However at high energies the growth of the total cross section guided by the Froissart bound \cite{Froissart} and by the behaviour of the observed data can be parametrized as a quadratic form in $\log (s/s_0)$ such as $\sigma \sim \log^2 (s/s_0)$, where we consider $s_0=1$ GeV, and we omit $s_0$ for simplicity.  The form of the differential cross section depends on specific assumptions  for  the real and imaginary  amplitudes, controled by dispersion relations (DR). In what follows we use the variable $E=(s-u)/4\,m$ as a crossing symmetric variable, which coincides with  the energy in the laboratory frame for $t=0$.

In another treatment for  very large energies in the forward region, if the real amplitude is neglected compared with the imaginary part, the scattering amplitude $F^N(E,t)$ is suggested to follow a scaling dependence \cite{Deus}, with $F^N(E,t)/F^N(E,0)=f(\tau)$ where $\tau$ is a combination of $E$ and $t$ variables. Using the $\log^2(E)$ dependence combined with the scaling function $f(\tau)$ the scattering amplitude is then written with the form,
\begin{eqnarray}
F^N(E,t)\sim  i\,C \,E\, \log^2(E) f(\tau)~.
\label{amplitude_pomeron_2}
\end{eqnarray}
  The bounds and constraints  of $f(\tau)$ were formally studied long ago \cite{Deus} in the context of axiomatic field theory, giving $f(\tau)\leq \kappa \exp(\sqrt{|\tau|})$, where $\kappa$ is constant and  the scaling variable is $\tau = t\log^2 E$.
The cross section corresponding to Eq.(\ref{amplitude_pomeron_2}) is not invariant under the transformation, $E \rightarrow -E$ (crossing symmetry).
  Following A. Martin \cite{Martin_Real}, in order to define a complex crossing symmetric function,  Eq.(\ref{amplitude_pomeron_2}) can be modified to 
\begin{eqnarray}
F^N(E,t)\sim  i\,C\, E\,\Big(\log (E)-i\frac{\pi}{2}\Big)^2 f(\tau')~.
\label{amplitude_pomeron_3}
\end{eqnarray}
The extension of $\log (E)$ in the complex plane changes the scaling variable to a complex variable
$\tau'=t\,[\log (E)-i\pi/2]^2$ which transforms the scaling function $f(\tau')$ in a complex quantity.

 However, it is not obvious that at large  energies these amplitudes, with both crossing and scaling, are well satisfied. Attempts have been made in this direction to analyse ISR energies with $f(s,t)=\sum_{i=1}^n\,a_i\,e^{b_i\,t}$ where $a_i$ and $b_i$ are free parameters  to be fitted at each energy  \cite{MenonFagundes}. To test the ideas of crossing and scaling we propose in the current work a modified model for the forward scattering at high energies using the analytical crossing symmetric forms explored by Block and Cahn \cite{Block:1985} and used in the COMPETE parametrization \cite{COMPETE} \begin{eqnarray}
&& \mathcal{R}_+(E,\omega)=(m+E)^{\omega}+(m-E)^{\omega}  ~\to ~\mathcal{R}_+(E,\omega)\simeq 2\,E^{\omega}\cos(\omega\,\beta)e^{-i\,\omega\,\beta} \nonumber \\
&&\mathcal{C}_+(E)=-((m+E)(m-E))^{1/2}  ~~~~~~\to ~~~ \mathcal{C}_+(E)\simeq i\,E \nonumber \\ 
&& \mathcal{L}_+(E)=\frac{1}{2}[\log(\frac{m-E}{E_0})+\log(\frac{m+E}{E_0})] ~~\to ~~ \mathcal{L}_+(E)\simeq\log(E)-i\,\beta \nonumber\\
&& \mathcal{R}_-(E,\omega)=(m+E)^{\omega}-(m-E)^{\omega} ~\to ~\mathcal{R}_-(E,\omega)\simeq 2\,i\,E^{\omega}\sin(\omega\,\beta)e^{-i\,\omega\,\beta} ~.\nonumber \\
 \label{analytical_forms}
 \end{eqnarray} 
It is easy to check that the LHS of Eqs.(\ref{analytical_forms}) are analytical by the interchange of $E\to-E$ keeping the forms crossing symmetric. Since we are interested in a high energy region (starting from ISR energies) the mass term on the LHS of Eqs.(\ref{analytical_forms})  can be safely neglected and the RHS approximation is justified. Crossing symmetry imposes  $\beta=\pi/2$, but here we let $\beta$ free for a test against the data. The subscripts (+) and (-)   refer to even and odd terms under crossing symmetry. We  generalize  Eq.(\ref{amplitude_pomeron_3}) writing
  \begin{eqnarray}
 F_{\mp}^N(E,t)= \mathcal{C}_+(E)\,\Big[P'+P_1'\,\mathcal{L}_+(E)+H'\,(\mathcal{L}_+(E))^2+R_1'\,\mathcal{R}_+(E,-\eta_1)\mp\,R_2'\,\mathcal{R}_-(E,-\eta_2)\Big]\, f(\tau')~, 
\label{amplitude_pomeron_4}
\end{eqnarray}
where $\mp$ are refereed to pp/p\=p respectively. The parameters $P'$, $P_1'$, $H'$, $R_1'$, $R_2'$, $\eta_1'$ and $\eta_2'$ are fixed numbers for all energies.

The $t$ dependence is embedded in the complex scaling variable $\tau'$ defined as
\begin{eqnarray}
\tau'(E,t)=\Big[b_0'+b_1'\,\mathcal{L}_+(E) +b_2'\,(\mathcal{L}_+(E))^2+b_3'\,\mathcal{R}_+(E,-\eta_3)\Big]\,t ~,
\label{tauL}
\end{eqnarray}
with $b_0'$, $b_1'$, $b_2'$, $b_3'$  as fixed quantities, while $\eta_3'$ is allowed to have energy dependence. We assume a  complex scaling function
\begin{eqnarray}
f(\tau')\equiv e^{\tau'}=e^{\Omega_R'(E,t)+i\,\Omega_I'(E,t)}~.%
\label{damping_f}
\end{eqnarray}
We can also approximate $s\approx 2\,m\,E$ and we re-write the above amplitudes using the $s$ variable keeping the same analytical dependence
  \begin{eqnarray}
 F_{\mp}^N(s,t)= \Big[F_{\mp}^{R}(s)+i\,F_{\mp}^{I}(s)\Big]\,f(\tau') = \Big[F_{\mp}^{R}(s)+i\,F_{\mp}^{I}(s)\Big]\, e^{\Omega_R(s,t)+i\,\Omega_I(s,t)}~.
\label{amplitude_pomeron_5}
\end{eqnarray}
Separating the real and imaginary parts we have
\begin{eqnarray}
F_{\mp}^{R}(s)=s\,\Big[\beta\,\Big(P_1+2\,H\,\log(s)\Big)-R_{1}s^{-\eta_1}\sin(\eta_1\,\beta)\mp R_{2}s^{-\eta_2}\cos(\eta_2\,\beta)\Big]~,
\label{TR01}
\end{eqnarray}
\begin{eqnarray}
F_{\mp}^{I}(s)= s\,\Big[P+P_1\,\log(s)+H\Big(\log^2(s)-\beta^2\Big)+R_{1}s^{-\eta_1}\cos(\eta_1\,\beta)\pm R_{2}s^{-\eta_2}\sin(\eta_2\,\beta)\Big]~, 
\label{TI01}
\end{eqnarray}
and 
\begin{eqnarray}
\Omega_R(s,t)=\Big[b_0+b_1\log(s)+b_2\Big(\log^2(s)-\beta^2\Big)+b_3\,s^{-\eta_3}\cos(\eta_3\,\beta)\Big]\, t~,
\label{REAL_SLOPE}
\end{eqnarray}
\begin{eqnarray}
\Omega_I(s,t)=-\Big[b_1\,\beta+2\,b_2\,\beta\,\log(s)-b_3\,s^{-\eta_3}\,\sin(\eta_3\,\beta)\Big]\, t ~.
\label{IMAG_SLOPE}
\end{eqnarray}
We adopt the suggestion from COMPETE parametrization \cite{COMPETE} for the amplitudes at $t=0$ with the same given parameters $P$, $P_1$, $H$, $R_1$, $R_2$, $\eta_1$ and $\eta_2$, while $\beta$ is left as a free parameter to test the crossing symmetry. The parameters 
that dictate the forward $t$ dependence of the differential cross section are $b_0$, $b_1$, $b_2$, $b_3$ (energy independent) while the energy dependence of  $\eta_3$ is obtained from differential cross section data.

The separation of the real and imaginary gives
\begin{eqnarray}
F_{\mp}^{R}(s,t)=  
F_{\mp}^{I}(s)\,\Bigg[-\sin \Omega_I(s,t)+\frac{F_{\mp}^{R}(s)}{F_{\mp}^{I}(s)}\,\cos \Omega_I(s,t)\Bigg]\,e^{\Omega_R(s,t)}
\label{TR1}
\end{eqnarray}
and
\begin{eqnarray}
F_{\mp}^{I}(s,t)=  
F_{\mp}^{I}(s)\,\Bigg[\cos \Omega_I(s,t) +\frac{F_{\mp}^{R}(s)}{F_{\mp}^{I}(s)}\,\sin \Omega_I(s,t) \Bigg]\,e^{\Omega_R(s,t)}~.
\label{TI1}
\end{eqnarray}
 The complex amplitude can be written in the compact form
 \begin{eqnarray}
\begin{pmatrix} 
F_{\mp}^{R}(s,t) \\ F_{\mp}^{I}(s,t)
\end{pmatrix}
=
s\,\sigma_{\mp}(s)\,\begin{pmatrix} 
\cos \Omega_I(s,t) & -\sin \Omega_I(s,t) \\
\sin \Omega_I(s,t)\, & \cos \Omega_I(s,t)  
\end{pmatrix}
\begin{pmatrix} 
\rho_{\mp}(s) \\ 1
\end{pmatrix}\,e^{\Omega_R(s,t)}~,
 \label{COMPLEX-FUNC}
 \end{eqnarray}
 where $\Omega_I(s,t)$ is the forward mixing angle between  $\rho_{\mp}$ and the unity. 
 The total cross sections for pp/p\=p are given by the optical theorem
\begin{equation}
\sigma_{\mp}(s) = \frac{F_{\mp}^{I}(s,0)}{s}~,
\label{sigma}
\end{equation}
and the the ratios of the amplitudes at $-t=0$ are
\begin{equation}
\rho_{\mp}(s) = \frac{F_{\mp}^{R}(s,0)}{F_{\mp}^I(s,0)}~.
\label{rho}
\end{equation}

Note that Eq.(\ref{COMPLEX-FUNC}) has trigonometric functions which make the sum of the absolute square of the real and imaginary nuclear parts as a simple exponential function 
\begin{eqnarray}
|F_{\mp}^{R}(s,t)|^2+|F_{\mp}^{I}(s,t)|^2= s^2\,\sigma_{\mp}^2(\rho_{\mp}^2+1)\,\,e^{2\,\Omega_R(s,t)}~,
\end{eqnarray}
 which does not  describe well the forward scattering data. To improve we add to the real part a shape function \begin{eqnarray}
G_{\mp}^R(s,t)=\sigma_{\mp}\,\frac{s\,t}{\Lambda^2-t}\,e^{\Omega_R(s,t)}~,
\label{shape-R}
\end{eqnarray} 
which is zero for $t=0$ and does not affect crossing symmetry.
The amplitude is then written as
  \begin{eqnarray}
F_{\mp}^N(s,t)\to F_{\mp}^N(s,t)+G_{\mp}^R(s,t)~
.
\label{amplitude_complete}
\end{eqnarray}
The sum of the squared amplitudes is
\begin{eqnarray}
|F_{\mp}^R(s,t)|^2+|F_{\mp}^I(s,t)|^2\simeq s^2\,\sigma_{\mp}^2(\rho_{\mp}^2+1)\,e^{2\,\Omega_R(s,t)}+2\,s\,\,\sigma_{\mp}\,G_{\mp}^R(s,t)\,(\rho_{\mp}\,\cos\Omega_I-\sin\Omega_I)\,e^{\Omega_R}+[G_{\mp}^R(s,t)]^2~.
\end{eqnarray} 
A more complete $t$ dependence of $G_{\mp}^R(s,t)$ may also contain terms with negative parity, for example, the Odderon (for a review see \cite{Carlo}) which may be important for larger $t$ values, specially at the dip region.  

The real part is approximatly linear in $t$ for small $t$ and accounts for a zero with an energy dependence, which corresponds to Martin's zero \cite{Martin}, while in the imaginary amplitude there is no zero, since this model does not intend to describe the dip region. This can be seen due to the lack of the minus sign in the rotation matrix Eq.(\ref{COMPLEX-FUNC}).

The derivatives of the logarithm of the real and imaginary amplitudes at $|t|=0$  are respectively,
\begin{eqnarray}
\frac{\partial}{\partial t}\log F_{\mp}^R(s,t)\Big|_{|t|=0} \simeq  -\frac{1}{\rho_{\mp}}\,\frac{\partial\,\Omega_I(s,t)}{\partial\, t}\Big|_{|t|=0}+ \frac{\partial\,\Omega_R(s,t)}{\partial\, t}\Big|_{|t|=0}+\frac{\partial\log(G_{\mp}^R(s,t))}{\partial t}\Big|_{|t|=0} \equiv \frac{B_{\mp}^R(s)}{2}~
\label{BR}
\end{eqnarray}
and
\begin{equation}
\frac{\partial}{\partial t}\log F_{\mp}^I(s,t)\Big|_{|t|=0} \simeq \rho_{\mp}\,\frac{\partial\,\Omega_I(s,t)}{\partial\, t}\Big|_{|t|=0}+ \frac{\partial\,\Omega_R(s,t)}{\partial\, t}\Big|_{|t|=0} \equiv \frac{B_{\mp}^I(s)}{2}~,
\label{BI}
\end{equation}
where $B_{\mp}^R$ and $B_{\mp}^I$ are refereed to as the effective slopes of the real and imaginary  amplitudes.
These derivatives determine the average slope of the differential cross section at $|t|=0$, which in terms of Eqs. (\ref{rho}), (\ref{BR}) and (\ref{BI}) is given by
\begin{eqnarray}
B_{\mp}(s)\equiv\frac{d}{d t}\log\Big(\frac{d\sigma_{\mp}}{dt}\Big)\Big|_{|t|=0} =   \frac{\rho_{\mp}^2B_{\mp}^R+B_{\mp}^I}{\rho_{\mp}^2+1}=2\,\frac{\partial\, \Omega_R(s,t)}{\partial t}\Big|_{|t|=0}+\frac{\rho}{\rho^2+1}\frac{1}{\Lambda^2}~. 
\label{slope}
\end{eqnarray}

The high energy behaviour ($\log^2s$) of pp and p\=p total cross section implies  $\rho \sim \pi/\log(s)$ for high energies \cite{Kinoshita}, which compared with Eq.(\ref{rho}) automatically imposes $\beta=\pi/2$. The exact forms of dispersion relations can be computed for the amplitudes and their derivatives according to our recent mathematical results \cite{DDR_EXACT}. However, the energy dependence of the parameter $\eta_3$ needs to be studied in detail.

We fix the parameters $P=36$ mb, $P_1=-1.5$ mb, $R_1\cos(\eta_1\,\beta)=36.61$ mb, $R_2\sin(\eta_2\beta)=26.15$ mb, $\eta_1=0.4473$ and $\eta_2=0.5486$, according to PDG parametrization (with some flexibility in the numerical values), living $H$ as a free parameter. The parameters $b_0=11.36$ GeV$^{-2}$, $b_1=0.058$ GeV$^{-2}$, $b_2=0.0079$ GeV$^{-2}$ and $b_3=-17.58$ GeV$^{-2}$ are fixed according to the global fit of all data analysed. The parameter $\beta$ responsible for the crossing symmetry is stable in most of the studied energies and is compatible with $\pi/2$. We leave $\eta_3$ as a free parameter.  We fit the experimental data from FNAL \cite{Kuz}, ISR \cite{Amaldi,Amos}, Sp\=pS\cite{Augier}, E710\cite{E710} and LHC \cite{A7,A8,T7,T8,T13}, covering a range of $\sqrt{s}$ from 23 GeV to 13 TeV, using 
\begin{eqnarray}   
\frac{d\sigma}{dt}=\frac{1}{16\pi(\hbar c)^2\, s^2} |F(s,t)|^2  ~   ,
\end{eqnarray}
where the amplitude $F(s,t)$ is 
\begin{eqnarray}
\frac{F(s,t)}{4\,\pi\,(\hbar c)^2\,s}=\frac{F_{\mp}^N(s,t)}{4\,\pi\,(\hbar c)^2\,s}+\,F_C(t)\,e^{i\Phi(s,t)}~.
\label{full-ampl}
\end{eqnarray}
The Coulomb interaction together with the proton form factor is written for pp and p\=p respectively
\begin{eqnarray}
F_C(t)=\mp \frac{2\,\alpha}{t}\Big(\frac{\Lambda^2}{\Lambda^2-t}\Big)^2~,
\label{Coulomb}
\end{eqnarray}
with $\alpha=1/137$ (the fine structure constant) and $\Lambda^2=0.71$ GeV$^2$ is the electromagnetic form factor scale. 

In spite of many attempts to calculate the relative Coulomb phase $\Phi$, it is still a open problem. The most used approaches such as West-Yennie, Slov'ev, Cahn and Kundrat \cite{phases}, were derived based on strong assumptions. First the WY approach was based in the Feynman diagrams considering the strong interactions as an effective vertex (similarly with Slov'ev). The electromagnetic Feynman diagrams were expanded assuming the proton as a point like-particle. On the other hand Cahn's approach which was based on the additivity of the eikonal functions for the Coulomb and strong interactions is not more than a guess. The two interactions have no reason the be additive in terms of eikonals. Besides the electromagnetic form factor is imposed by hand. In this sense, as explained in our previous
work the relative Coulomb phase could be tested as zero without problems \cite{us}. 

\bigskip

\section{COMPARISON WITH DATA}

\bigskip

The datasets have been  analysed in limited $t$-range ($0< |t|< 0.2$) GeV$^2$ chosen appropriately   in order to account for the stability of  $\beta$, which in our analysis is manifestly positive. The positiveness of the real amplitude for $|t|$ near zero was recently proved by A. Martin and T. T. Wu \cite{Martin_2017} and this supports our results for $\beta$.

We obtain that all datasets analysed are well represented by our model with reasonable small $\chi^2/ndf$.  The fits show that  $\beta$ is compatible with $\pi/2$ for all datasets. This parameter is related with the phase of the complex nuclear amplitude. Thus, if we consider $\beta=\pi/2$ as an input in Eq.(\ref{amplitude_pomeron_4}) the scattering amplitude becomes crossing symmetric under $s\to u$ for fixed $-t$. 

\begin{figure}
\includegraphics[scale=0.40]{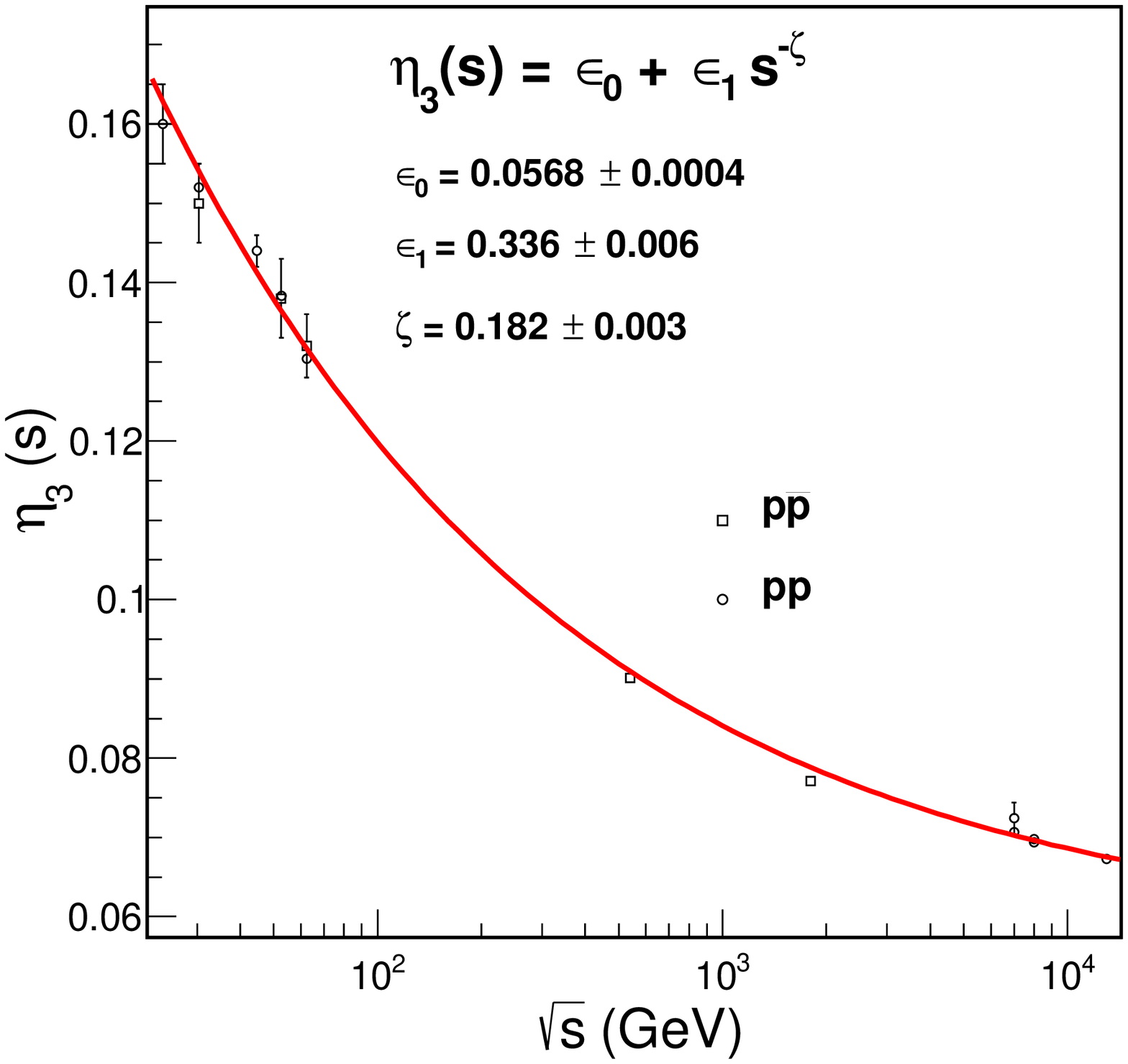}\includegraphics[scale=0.40]{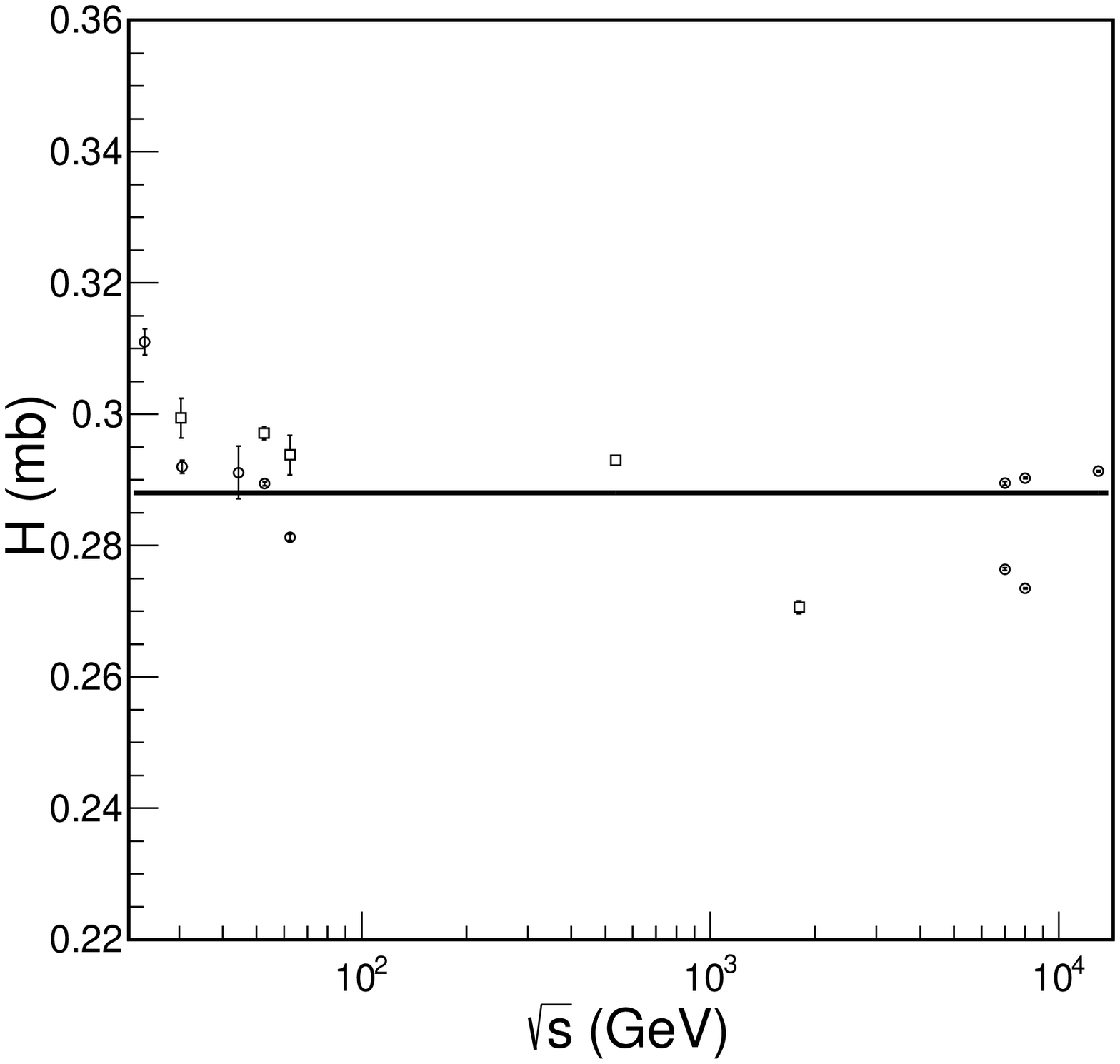}
\caption{The LHS plot shows the  energy dependence of the parameter $\eta_3$.  For high energies (LHC and beyond) the value of the parameter  tends to stay constant around $\epsilon_0$. The RHS figure shows the behaviour of the parameter H which is approximately constant in the large range of the analyzed energies. The numerical value is similar to the one proposed by COMPETE parametrization.}
\label{H-eta3}
\end{figure}

The energy dependence of $\eta_3(s)$, shown in the LHS of Fig.\ref{H-eta3}, can be parametrized with a power of $s$   as 
\begin{eqnarray}
\eta_3(s)=\epsilon_0+\epsilon_1\,s^{-\zeta}~,
\label{eta-3}
\end{eqnarray}
with the parameters  $\epsilon_0=0.0568\pm 0.0004$, $\epsilon_1=0.336\pm 0.006$ and $\zeta=0.182\pm 0.003$.
 These parameters seem to be constant with energy, supporting the idea of the existence of a complex scaling function $f(\tau')$. 
  The regular dependence of $\eta_3(s)$ is remarkable, and for very large energies this quantity tends to stay constant around $\epsilon=0.056$. The RHS of Fig.\ref{H-eta3} shows the values of H obtained for the analysed energies. We see that the numerical values fluctuates according to different normalizations of the experiments, but on average it stays constant. In this sense in order to make some predictions we fix the average value of $H=0.288$ mb.

 In Table \ref{predictions}  we show the obtained parameters $H$, $\eta_3$ for all analysed energies and the extracted forward quantities $\sigma$, $\rho$, $B$, $\sigma_{\rm elas}$. We also show predictions for the energies  0.2, 0.9, 2.76, 14 and 57 TeV.

\begin{figure}
\includegraphics[scale=0.40]{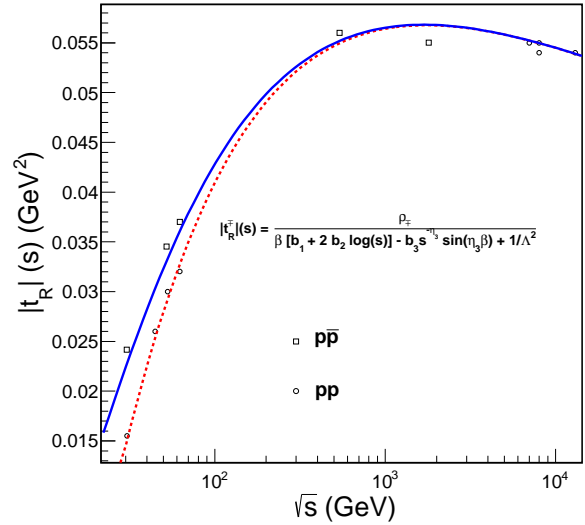}
\caption{The figure shows the zero of Martin for the real part obtained from our model. The symbols represent the zeros obtained from the fits. The curves are obtained with $\eta_3(s)$ and $|t_R|(s)$ from Eqs.(\ref{eta-3}) and(\ref{tR0}).
}
\label{martin}
\end{figure}

An important feature of the model is Martin's zero in the real part of pp and p\=p \cite{Martin}. Setting Eq.(\ref{TR1}) equals zero we obtain the solution for Martin's zero. However, since we are working in very small $-t$ the trigonometric and the shape functions can be expanded in powers of $t$.  We take the leading terms in the real amplitude and we write the position of the zero
\begin{eqnarray}
|t_R|\simeq\frac{\rho_{\mp}(s)}{\beta\,[b_1+2\,b_ 2\,\log(s)]-b_3\,s^{-\eta_3(s)}\sin[\eta_3(s)\beta]+1/\Lambda^2}~.
\label{tR0}
\end{eqnarray}
Fig.\ref{martin} shows the zeros of Martin for pp and p\=p. For ISR energies the position of the zero moves to higher $-t$, but at TeV energies it starts to decrease as predicted by the theorem of Martin\cite{Martin}.

\begin{figure}
\includegraphics[scale=0.4]{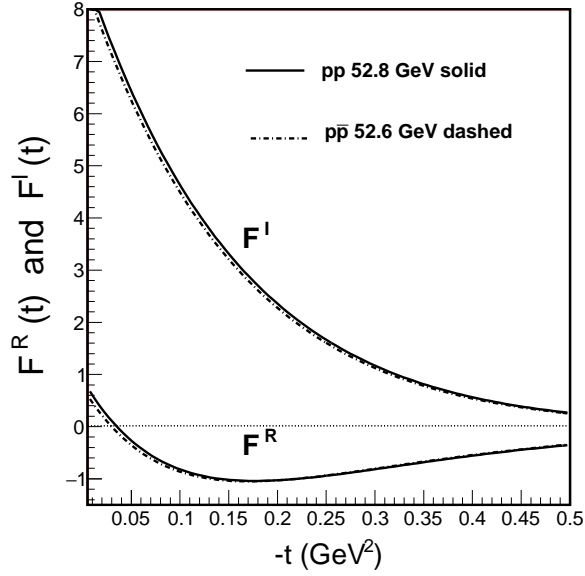}
\caption{The  figure shows the real and imaginary amplitudes for pp and p\=p at $\sqrt{s}=$52 GeV. The real part has a zero close to $|t|\sim$0.034 GeV$^2$. It is important to stress that these amplitudes are realistic until $-t\simeq$0.2 GeV$^2$. Beyond this range, other terms might be necessary in the real and imaginary parts. 
In this figure we only consider nuclear part.}
\label{pp-52-ampl}
\end{figure}

Fig.\ref{pp-52-ampl} shows the  amplitudes for pp and p\=p at similar energies about (52 GeV). The real parts are positive at $-t=0$ and have zeros near the origin.  The magnitude of the imaginary part is much larger than the real part for low $-t$ values. 

\begin{figure}
\includegraphics[scale=0.4]{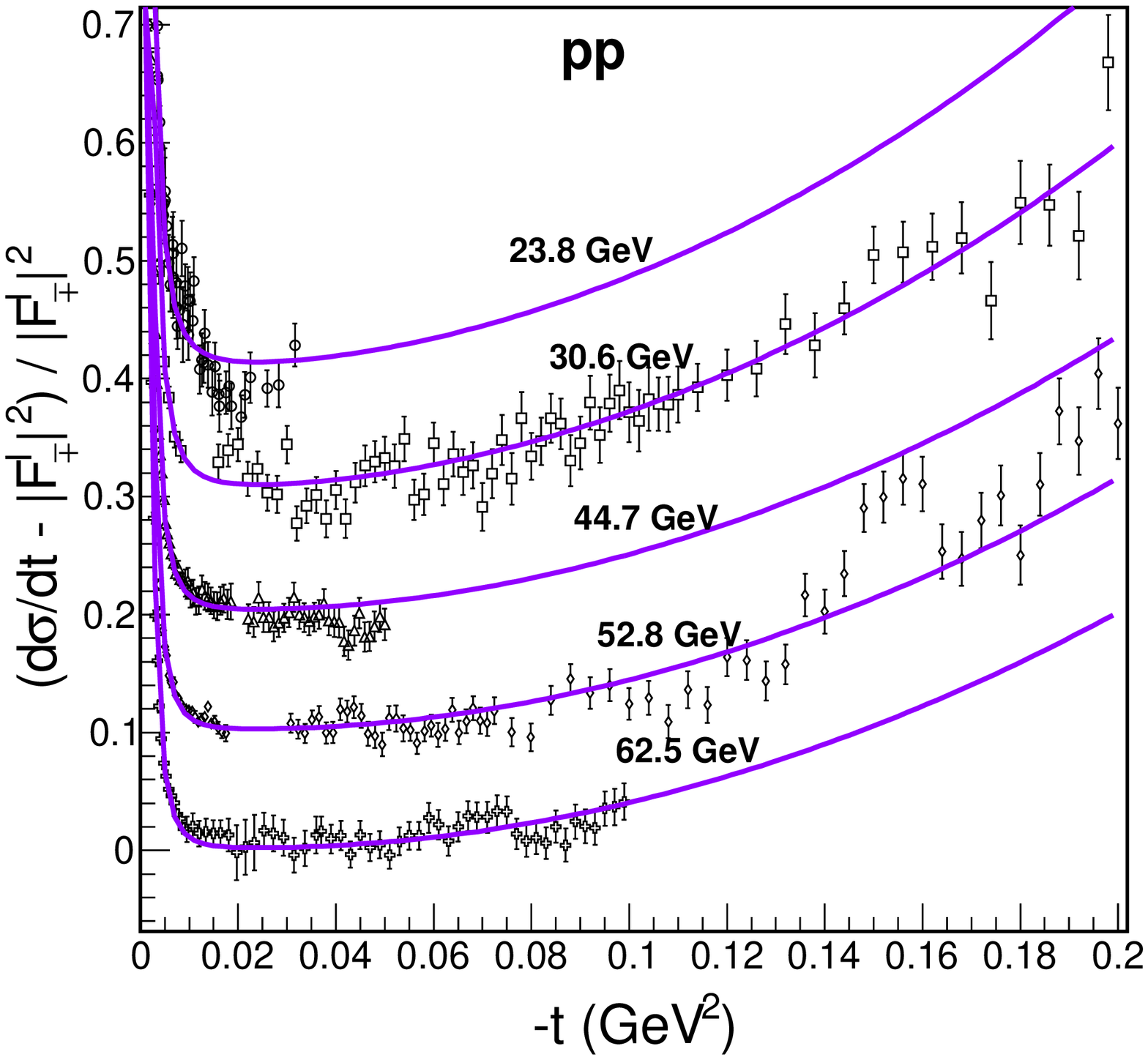}
\includegraphics[scale=0.4]{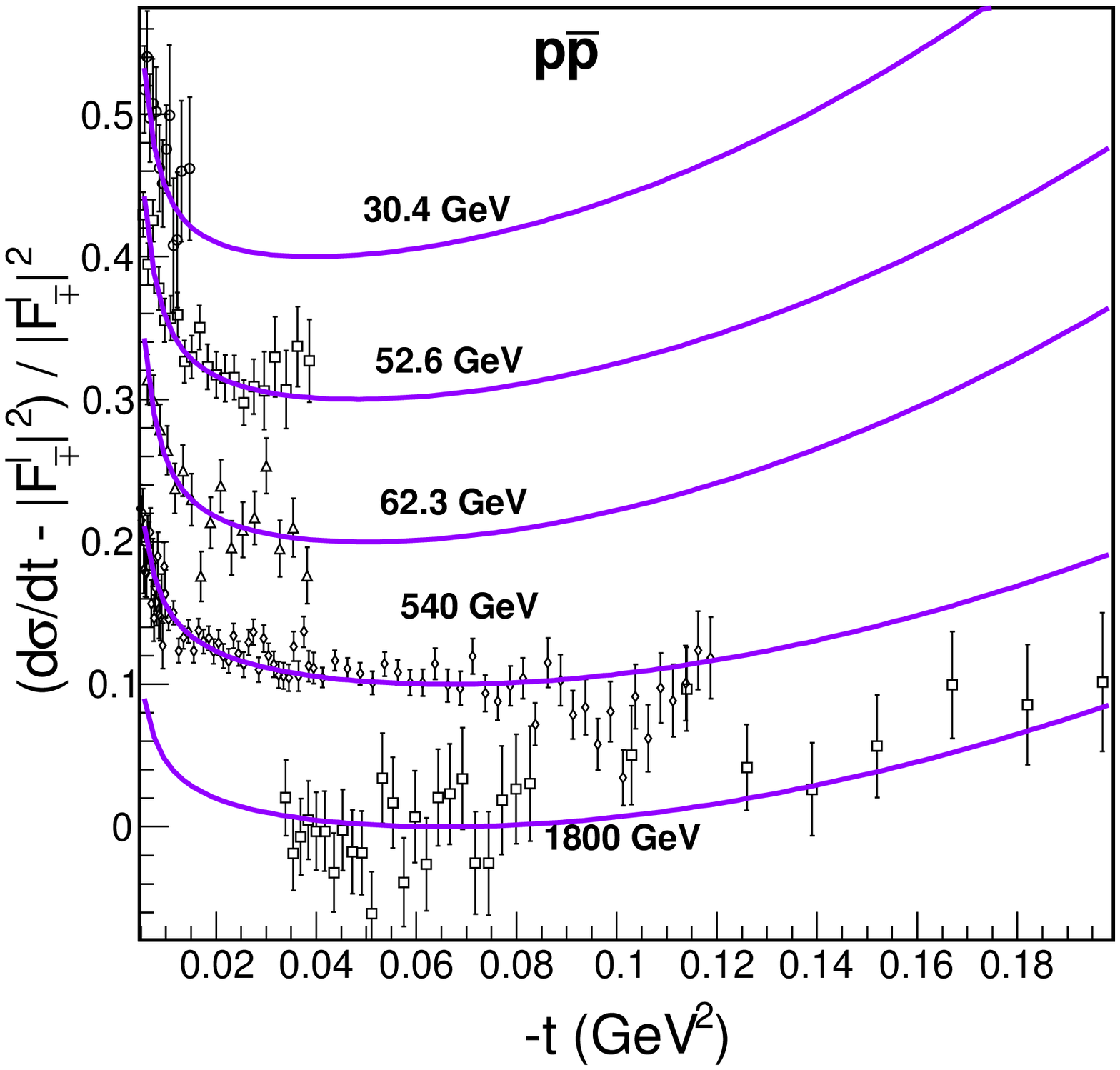}
\caption{The LHS plot shows the quantity $\Big(d\sigma/dt-\pi(\hbar c^2) |F_{\mp}^{I}|^2\Big)/\Big(\pi(\hbar c)^2 |F_{\mp}^{I}|^2\Big)$ for pp at ISR energies. The subtraction of the squared of the imaginary part essentially removes the pure exponential dependence of the differential cross section, putting in evidence the non-exponential behaviour of the distributions. To avoid the piling of the data at different energies we add to the $y$ axis multiples of 0.1. The RHS plot is similar, for p\=p data from ISR/Cern and E710/Fermilab. Again we add to the curves and the data multiples of 0.1.}
\label{LOW-ENERGIES}
\end{figure}

Fig.\ref{LOW-ENERGIES}  shows the ratio $\Big(d\sigma/dt-\pi(\hbar c^2) |F_{\mp}^{I}|^2\Big)/\Big(\pi(\hbar c)^2 |F_{\mp}^{I}|^2\Big)$  for $\sqrt{s}=$ 23, 44, 52, 62, 540, 1800 GeV for both pp and p\=p. This ratio removes from the differential cross section the simple exponential behaviour, which  is essentially dictated by the imaginary part, allowing to investigate with detail the structure of the real amplitude.
We observe that the curvature at all energies is created by the zero of the real part together with and the magnitude of the real slope.

\begin{figure}
\includegraphics[scale=0.42]
{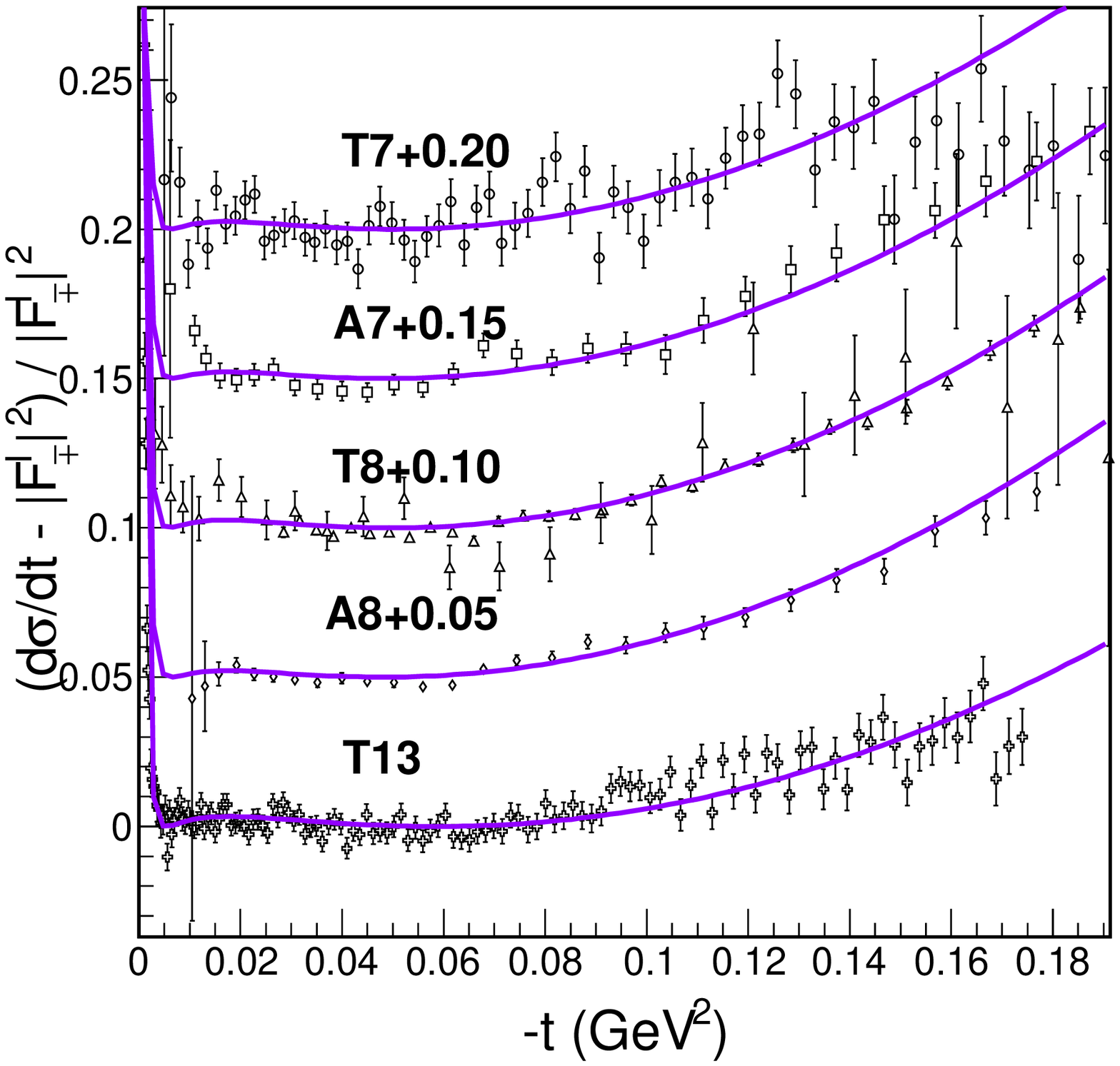}
\includegraphics[scale=0.4]{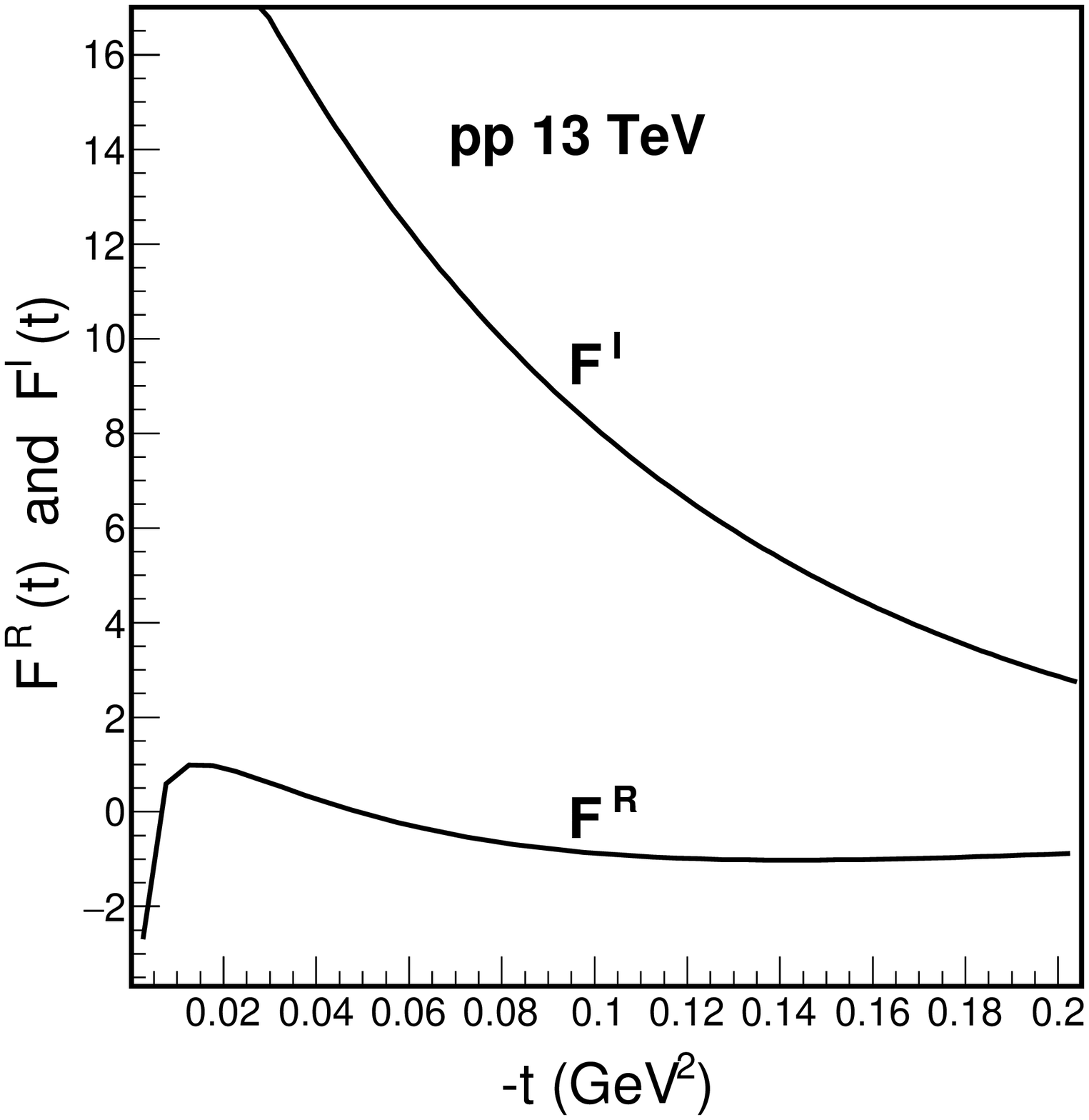}
\caption{The LHS plot shows $\Big(d\sigma/dt-\pi(\hbar c^2) |F_{\mp}^{I}|^2\Big)/\Big(\pi(\hbar c)^2 |F_{\mp}^{I}|^2\Big)$ at LHC energies. As before we add multiples of 0.1 in the curves and data to separate them. Note that at these energies a  'soft peak' is presented near $|t|=0.006$ GeV$^2$. The RHS plot shows the real and imaginary amplitudes at 13 TeV together with the Coulomb interaction. Since the Coulomb amplitude is negative for pp and the real nuclear part is positive for small $-t$ there is a region where their sum cancels out, producing a minimum. This effect explains the peak that could be present for very forward scattering at high energies.}
\label{LHC}
\end{figure}

The LHS of Fig. \ref{LHC} shows the ratio $\Big(d\sigma/dt-\pi(\hbar c^2) |F_{\mp}^{I}|^2\Big)/\Big(\pi(\hbar c)^2 |F_{\mp}^{I}|^2\Big)$  for LHC energies $\sqrt{s}=$7, 8 and 13 TeV. As in the previous figures involving the data we add factors multiples of 0.05 to separate the data. For these LHC energies we note  that a 'soft peak' structure is presented in the very forward region. In the RHS figure we show the real and imaginary amplitudes at 13 TeV, but this time we include the Coulomb amplitude. For pp scattering the Coulomb amplitude is negative while the real nuclear part is positive. The interplay between them produces a zero in the very forward range which can be seen in the figure.

\begin{table}[t]
\begin{center}
 \vspace{0.5cm}
\begin{tabular}{ccc|cccc|cc}
\hline
\hline 
            &      \multicolumn{2}{c|}{parameters}       & \multicolumn{4}{c|}{derived quantities}   & &\\
  $\sqrt{s}$ & $H$  & $\eta_3$  &  $\sigma$ & $\rho$& $ B$   & $\sigma_{\rm elas.}$ &$\chi^2/ndf$  & Refs \\
     (GeV)  &   (mb)   &    &   (mb) &   &  (GeV$^{-2})$& (mb) & &\\
      \hline
  \hline
  pp   &  & &           &      &  & &   &   \\
    \hline
   23.882    & 0.311$\pm$0.002 & 0.16 (fix)  &  39.57  & 0.034    & 11.77  & 7.27 & 90.7/62 & \cite{Kuz}\\

  \hline
   30.6    & 0.292$\pm$0.001 & 0.1522$\pm$0.001  &  39.79  & 0.049    & 12.23 & 7.03 & 94.1/68 &\cite{Amaldi}\\
  \hline
   44.7    & 0.291$\pm$0.0004  & 0.144 (fix)  & 41.51  & 0.077     & 13.12  & 7.08 & 87.3/67& \cite{Amaldi} \\
 \hline
   52.8    & 0.2894$\pm$0.0003 &  0.1383$\pm$0.0003    &  42.41  & 0.088     & 13.26  & 7.30 & 245/88 &\cite{Amaldi}\\
 \hline
   62.5    & 0.2812$\pm$0.0005  & 0.1304$\pm$0.0004    &  42.76  & 0.092      & 13.15 & 7.49 & 111.1/62 &\cite{Amaldi}\\   
    \hline
   200*   & 0.2887 (fix) & 0.106 (fix)    &  52.05 & 0.133      & 14.61  & 9.94 & -\\   
       \hline
   900*   & 0.2887 (fix)  & 0.085 (fix)    &  68.38 & 0.145      & 16.43  & 15.15 &- \\   
    \hline
   2760*      & 0.2887 (fix)  &  0.076 (fix)   & 84.04 & 0.143     & 18.23 & 20.51 & -\\
       \hline
   7000      & 0.2895$\pm$0.0003  &  0.0735$\pm$0.0002   & 99.51  & 0.138     & 20.39& 25.57 & 74.4/59 &\cite{T7} \\
          \hline
   7000      & 0.2764$\pm$0.0002  &  0.0707$\pm$0.0001   & 95.43  & 0.136     & 19.90 & 24.11 & 42.9/33 &\cite{A7} \\
    \hline
   8000      & 0.2903$\pm$0.0001  &  0.0694$\pm$0.0001   &  102.12  & 0.137  & 19.65 & 27.99 & 72.5/58 &\cite{T8} \\
    \hline
   8000      & 0.2735$\pm$0.0001  &  0.0698$\pm$0.0001   &  96.74  & 0.135 & 20.11 & 24.51 & 28.8/25 &\cite{A8}\\
 \hline
   13000      & 0.2913$\pm$0.0001  &  0.0673$\pm$0.0001   &  111.43  & 0.134     & 20.99 & 31.11 & 149.4/126 & \cite{T13}\\
\hline
   14000*      & 0.2887 (fix)      & 0.067 (fix)    & 112.65    &  0.132     &  21.13   & 31.15 & -\\
 \hline
    57000*      & 0.2887 (fix)     & 0.063 (fix)    & 141.59    &  0.123     &  24.19   &  42.93& -  \\
 \hline
  \hline
  p\=p   &   &     &      &      &   &   &    \\
  \hline
   30.4    & 0.2994$\pm$0.003 &  0.15 (fix)    &  41.32  & 0.076     & 12.05  & 7.73 & 22.4/25 &\cite{Amos}\\
\hline
   52.6    & 0.2971$\pm$0.001 & 0.138 (fix)    &  43.44  & 0.102     & 13.24  & 7.69 & 29.9/27 &\cite{Amos}\\
   \hline
   62.3    & 0.2938$\pm$0.003 & 0.132 $\pm$0.004    &  44.13  & 0.107     & 13.32  & 7.89 & 19.9/15 &\cite{Amos}\\
   \hline
   540   & 0.2930$\pm$0.0004 & 0.0901$\pm$0.0003    &  62.95  & 0.145  & 15.65  & 13.52 & 164.9/97 & \cite{Augier}\\
      \hline
   1800   & 0.2706$\pm$0.001  & 0.0771$\pm$0.0004    &  73.71  & 0.141  & 17.19  & 16.78 & 43.8/53 &\cite{E710}\\
 \hline
\\
 \end{tabular}
 \caption{ The left part of the table shows the parameters obtained in each analysed energy. It is interesting to note that $H$ is almost constant in all domain. The parameter $\eta_3$ tends to saturate at high energies, allowing the extrapolation for larger energies. The right part of the table shows the derived forward quantities. The values of $\sigma$, $\rho$ and $B$ are in accordance with the parameters given by the experimental papers.  The (*) are  predictions whose there no analyzable data. }   
\label{predictions}
\end{center}
\end{table}

The difficulties in the determination of the parameter $\rho$ from the data are well known for various phenomenological models. 
A proper determination depends on the analytical form used to parametrize the nuclear interaction and on the interference with the Coulomb interaction which is still an open problem. The quality of the experimental data in the interference region is crucial for this purpose. In the present model the  analytical connection between the real and imaginary parts  controls the fit instabilities, constraining $\rho$
from dispersion relations. In this sense we understand that the Odderon term is not necessary to explain the LHC forward data. Other analysis also consider only the even (Pomeron) term to explain the forward data \cite{Donachie}.

\section{Conclusion}

 In recent studies of the LHC data on pp elastic scattering with independent real and imaginary amplitudes \cite{us}, specific features of the real part, such as the position of the zero and the magnitude and sign of the amplitudes  were investigated, and the parameters were determined with high precision. In the present work we discuss properties of analyticity and crossing symmetry of the amplitudes in the forward regime. It is important to remark that the real and imaginary amplitudes Eq.(\ref{COMPLEX-FUNC}), when expanded to small $-t$ values have similar forms on $t$ compared with the amplitudes in previous work \cite{us}. The main difference is in the imaginary part of \cite{us} which contains an additional linear factor on $t$ that accounts for the existence of the dip in the differential cross section for larger $t$ values. This difference influences the parameters of the real part giving small modifications in the quantities such as $t_R$ and the effective slope $B_R^{\rm eff}$.  
 
  Advocated in \cite{Deus}, the geometric scaling  in hadronic collisions is achieved when the $s$ and $t$ dependence can be written in terms of a single variable $\tau(s,t)$. Until the LHC energies the experimental data on elastic pp and p\=p does not follow the geometric scaling. One way to observe this mismatch  is through the ratio between elastic and total cross section \cite{deDeus:2013} that should be constant with the energy and clearly this is not the case with the existing collider data. 
 Since in our model we can extrapolate our results to higher energies we can calculate the asymptotic ratio  $\sigma/\sigma_{\rm tot}\to 0.37$. However we are aware of the limitations of the model at very high energies and we do not claim that this is the final answer. Usually the scaling variable is defined to be a real function  $\tau(s,t)=-t\,\sigma_{\rm tot}(s)$, but this is an arbitrary choice. We define  a complex variable $\tau'$ such that the scaling behaviour is obeyed for all analysed energies, i.e., the ratio  $F^N(s,t)/F^N(s,0)=f(\tau')$ is function of a single variable starting already from ISR energies. Attempts have been made to extract universal properties of the position of the dip in the elastic differential cross section by the re-definition of the scaling variable \cite{Fagundes:2014} achieving relative success. 
In the microscopic level the geometric scaling was observed in Deep Inelastic Scattering for  $\sigma_{\gamma^{*}\,p}$ cross section \cite{Stasto}. In this process the scaling variable is given as  $\tau=Q^2/Q_s(x)$, which is a combination of the  virtuality of the photon $Q^2$ and the saturation scale for gluons $Q_s(x)$.  In principle there should exist a connection between the gluon saturation and the Froissart bound of the total hadronic cross sections. Works in this direction \cite{Froissart_Saturation} have been developed but so far, no conclusive answer was given for pp or p\=p processes.

The impact parameter space is interesting to investigate the geometrical features of the hadronic interactions and the matter distribution within the proton. However the impact parameter is not directly observed from the data and in order to obtain a complete profile function starting from the momentum space, one should be able to provide the complete $t$ dependence of the elastic amplitudes. Recently there have been some interest in the properties of the geometric space and some authors \cite{b-space} claim  the existence of a peripheral behaviour in the proton structure in the LHC energies called hollowness. In this description the inelastic profile functions at $b\approx 0$ is reduced as the energy increases meaning that the periphery of the proton is more active. It is important however to stress that the profile functions are not necessarily  genuine differential cross sections in b-space, but, since they behave monotonically and continuously in b, it is natural to relate these profile functions with the geometric dependence of the
interaction. In the present work since we are interested in the very forward region the corresponding b-space amplitudes are not completely developed. Once we incorporate the larger $t$ dependence in our model we will be able to discuss the physical properties of the b-space.

To summarize, in the present work we propose a phenomenological tool to interpolate and extrapolate the experimental data, obeying crossing symmetry, dispersion relations and allowing the existence of a geometric scaling at asymptotic energies. This is a very useful information since historically some experiments give different results at the same energy (see Totem and Atlas for pp at 8 GeV and Fermilab for p\=p at 1.8 GeV).  In addition we present precise predictions for the energies 200 and GeV which were recently released in the EDS Blois conference \cite{blois} giving the results $\sigma_{\rm tot}=51.81\pm 0.2$ mb, $\sigma_{\rm el.}=9.74\pm 0.02$ mb and $B=14.32\pm 0.09$ GeV$^{-2}$ and we see that those results are  predicted in the present work.  
  
\clearpage

 \section*{Acknowledgments}
The author thanks Erasmo Ferreira, Takeshi Kodama and Tri Nang Pham for useful discussions and reading of the manuscript. 
The author also thanks the Brazilian agency CAPES for financial support.

\end{document}